\documentclass[a4paper, 10pt, conference]{ieeeconf}      

\IEEEoverridecommandlockouts                              

\usepackage{graphics} 
\usepackage{epsfig} 
\usepackage{amsmath} 
\usepackage{xcolor}
\usepackage{subcaption}

\usepackage{xspace}
\usepackage{soul}
\usepackage{mathtools}
\usepackage{makecell}
\usepackage{nicefrac}
\usepackage{float}
\usepackage{wrapfig}

\definecolor{dgreen}{rgb}{0.0, 0.5, 0.0}

\usepackage{textcomp}

\makeatletter
\newcommand{\subalign}[1]{%
	\vcenter{%
		\Let@ \restore@math@cr \default@tag
		\baselineskip\fontdimen10 \scriptfont\tw@
		\advance\baselineskip\fontdimen12 \scriptfont\tw@
		\lineskip\thr@@\fontdimen8 \scriptfont\thr@@
		\lineskiplimit\lineskip
		\ialign{\hfil$\m@th\scriptstyle##$&$\m@th\scriptstyle{}##$\crcr
			#1\crcr
		}%
	}
}

\title{\LARGE \bf
Process Optimization of Black Soldier Fly Egg Production via Model Based Control*
}

\author{Alexander Kobelski$^{1}$, Arne-Jens Hempel$^{2}$, Murali Padmanabha$^{1}$, Luiz-Carlos Wille$^{1}$ and Stefan Streif$^{1}$
\thanks{*This research was funded by the German Ministry for Education and Research (BMBF) in the frame of the CUBES project, grant number 031B0733D.}
\thanks{$^{1}$With Faculty of Electrical Engineering and Information Technology, Automatic Control and System Dynamics, Technische Universitaet Chemnitz, 09111 Chemnitz, Germany
        {\tt\small \{alexander.kobelski,murali.padmanabha, stefan.streif\}@etit.tu-chemnitz.de}}%
\thanks{$^{2}$Arne-Jens Hempel is now with the Department of Digital Engineering, Staatliche Studienakademie Glauchau,
        08371 Glauchau, Germany
        {\tt\small arne-jens.hempel@ba-sachsen.de}}%
}

\begin{document}

\maketitle
\thispagestyle{empty}
\pagestyle{empty}

\begin{abstract}                
	Black soldier fly (BSF) larvae (\textit{Hermetia illucens}) are a valuable protein source for manufacturing animal feed.
	To maximize their production, both the quantity and quality of their reproductive cycle, i.e. egg production during oviposition, must be increased.
	In artificial environments, flies often sit idle in cages without mating, depleting their energy reserves and resulting in lower egg production per female.
	
	By controlling environmental conditions such as temperature and light inside breeding cages, the flies may be stimulated in a way that improves egg output.
	However, this stimulation increases the energy demand of the process and may stress the flies, resulting in reduced egg production.	
	Therefore, control must be applied in a careful way, which requires knowledge of the egg production cycle.
	
	In this work, a mathematical model describing the various fly life stages and their transition to the egg production process is developed.
	Relevant factors are identified and their effect on the fly life and egg production is mathematically described.
	Parameters are identified using data from literature and goodness of fit is evaluated.
	Using the model, an optimal control problem is formulated with the goal of minimizing the energy costs and increasing the egg production quantity. 
	In Simulation, our approach showed 13\% higher output in shorter time at reduced energy costs compared to a (standard) constant setpoint approach.
	Optimal control could reach same amount of egg in 60~\% of time compared to standard scenario.
\end{abstract}

\section{Introduction}
According to the FAO \cite{FAO_SOFA_2021}, recent crises have shown the vulnerability of agrifood systems to shocks and stress resulting in global food insecurity and malnutrition.
Robust food sources are needed for more resilient and reliable food supply.
Insect proteins from the Hermetia Illucens, also commonly known as black soldier fly (BSF), are a promising alternative to common protein sources like fish meal.
The BSF larvae can feed on a variety of substrates such as animal feed, algae, or waste \cite{Liland2017,Surendra2020}, which makes their rearing less dependent on global trade fluctuations. 
The dried larvae are rich in protein \cite{Spranghers2017} and can be used as supplements for animal rearing and aquaculture \cite{Liu2017}.
While there is a lot of work being done on the rearing process of the larvae \cite{Bava2019,Padmanabha2020,Padmanabha2022}, the reproduction processes of the matured flies have not yet received as much attention \cite{Lemke2022}.

For business, the production of eggs and young larvae often poses a bottleneck on the maximum rearing capacities of the production plant.
In standard practice, the flies will often sit idle in their cages without mating, delaying egg production and consuming their energy reserves which are required for high egg output.
The mating and oviposition process however can be influenced through various controllable variables, such as light, temperature, energy rich liquids or sex-ratio \cite{Zhang2010,Tomberlin2002b,Bertinetti2019,Hoc2019}.
The standard approach here is to choose static set points.
However, flies require different conditions for their life stages, e.g. stimulation by light only has an effect after flies reached sexual maturity.
In addition, excessive stimulation leads to constant movement and stress, which depletes the flies' energy reserves, resulting in a shortened life span and fewer eggs per female.
However, this in turn can speed up the process, i.e. a new reproductive cycle can be started more quickly.
Egg output, energy consumption, stress management, production speed --- optimizing this process requires a systematic control approach.

In this work a mathematical model for the purpose of process optimization of BSF egg production is presented.
First, the fly life cycle, the mating process as well as the egg production are explored and abstracted into mathematical models.
The influence of control factors such as temperature and light on the mating rate, life span and energy reserve utilization are also modeled.
The model parameters are identified and compared to data from literature.
A model based optimal control sequence for light and temperature is computed by the optimal controller developed and compared to a standard approach in simulations.
The experiment design of Nakamura et al. \cite{nakamura2016small} is taken as a benchmark in this work. 
Air temperature in the breeding cage is 25~$^\circ$C, flies are exposed to 16 hours of light per day and have access to sugar water.
Results are compared to each other and the benefits of optimal control in egg production are highlighted.

\section{Fly life cycle, mating, egg production and death}
This section introduces a mathematical model for the fly life cycle from emergence to adulthood while also considering oviposition processes.
Since the aim is to later control egg output using this model, environmental factors such as lighting and temperature are included.
Model equations are mostly formulated in a mechanistic way to allow for physical interpretation of equations.

BSF originates from South America and is adapted to warm and humid climates.
In nature, larvae feast as much possible and than pupate at warm and dry location.
After emergence, it takes about two days until they reach sexual maturity.
They cannot consume any food except for fluids like water or nectar \cite{Bruno2019}.
Their health and potential egg output is very dependent on their success during the larval stage.
In daylight, they will look for a partner who they mate with.
Approximately two days after mating, females oviposit and die shortly after.
Males can mate multiple times, depending on their remaining energy reserves \cite{Tomberlin2002b,Giunti2018,Hoc2019}.
This cycle has to be considered for the artificial reproduction setup.

The envisioned reproduction setup is a breeding cage with a certain number of male and female pupae $N_\mathrm{m}$ and $N_\mathrm{f}$ placed within.
In practice, the number of pupae placed in the cage will be big enough to assume that the number of male and female are the same.
For simplicity reasons, flies of both sexes are considered to only mate once.
Temperature and light are considered control variables, i.e. they can be changed to influence fly reproduction.
Flies will go through their life stages and mate with each other.
Female become fertilized and lay eggs.
Eggs are collected and taken out of the system, no new pupae or flies are introduced during a reproduction cycle.
Once population numbers are sufficiently low or too few eggs are layed per day ($\approx 10\textrm{--}14$ days) the cage is cleared of remaining flies and new pupae are introduced. 

The basic structure for the fly life cycle is a compartment model, i.e. life stage transition model, coupled with an energy balance.
Flies have transition rates proportional to population size from one life-stage to the next.
Only certain compartments have the ability to contribute to reproduction processes, others do not.
Additionally, flies have a constantly declining energy reserve pool -- low reserves result in lower egg output and higher chance of death.
Note that equations apply for both males and females, however parameters may differ.
For the sake of readability parameter indices in equations for male and female are omitted. 
If equations or parameters apply only to one sex it will be explicitly stated.

First, environment factors are introduced, then equations for the energy reserve dynamics are explained and lastly the life stage transition dynamics are described and combined with the influence and energy reserve models.

\subsection{Factors influencing the fly life cycle}
Recall that the aim is control and optimization of BSF egg production.
Consequently, influence factors have to be identified and their impact quantified.
A literature review revealed that the three most influential factors during the fly life are temperature, lighting and feed \cite{Holmes2012,Chia2018,Bertinetti2019,Hoc2019}.
Influence of relative air humidity has been studied for egg eclosion and adult emergence \cite{Holmes2012}, but data for the purpose of this model were not sufficient.

\begin{figure}[h]
	\centering
	\includegraphics[width=0.95\linewidth]{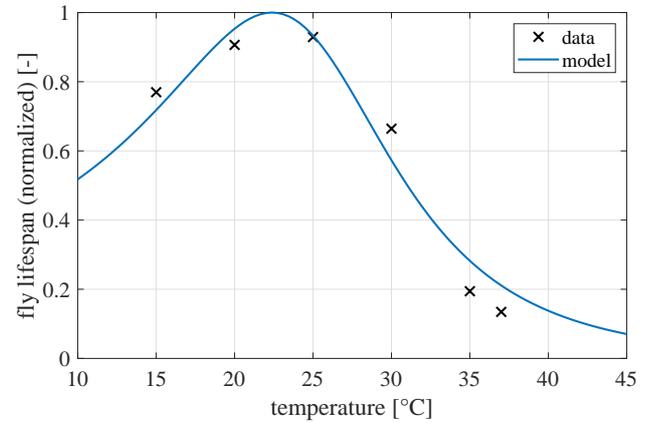}
	\caption{BSF adult fly lifespan at different temperature. Flies live longest at approximately $23^\circ \mathrm{C}$. Data is from \cite{Chia2018}.
	}
	\label{fig:lifespan}
\end{figure}

Temperature influences three properties of the fly life: lifespan, fertility (eggs per female) and life stage transition rates.
A modified Logan equation (Logan-10), see \cite{Logan1976} and \cite{Padmanabha2020}, is used to describe the influence of temperature 
\begin{align}
\label{eq:u_logan_main}
r_\mathrm{T}(T)&= \alpha[1+k_\mathrm{L}\exp(-p\cdot(T-T_\mathrm{R}))+\exp(-\vartheta)]^{-1}\\
\label{eq:logan_tau}
\vartheta&=\frac{T_\mathrm{let}-T}{\Delta T},
\end{align}
where $T_\mathrm{let}$ is the lethal maximum temperature, $T_\mathrm{R}$ is a reference temperature and $\Delta T$ is the width of the high temperature boundary layer.
Parameter $\alpha$ is the maximum observed process rate, $p$ is sensitivity of the rate to temperature changes and $k_\mathrm{L}=\dfrac{\alpha-k_\mathrm{base}}{k_\mathrm{base}}$ with $k_\mathrm{base}$ the minimum rate at lower temperature threshold.
Parameters were estimated using Matlabs \textit{lsqcurvefit} with data from \cite{Chia2018}, see Fig.~\ref{fig:lifespan} and Table~\ref{tab:parameters_T}.


Egg production per female in dependence of temperature is also modeled using a Logan equation with data from \cite{Chia2018}.
Parameters can be found in Table~\ref{tab:parameters_T}.

Temperature influence on transition rates $k$ between life stages is modeled in a similar way using another data set from \cite{Chia2018} for the time to first oviposition.
Parameters can be found in Table~\ref{tab:parameters_T}. 

\begin{table}[h]
	\centering	
	\begin{tabular}{|c||c|c|c|c|c|c|}	
		\hline 
		& $\alpha$ & $k_\mathrm{L}$ & $p$ & $T_\mathrm{let}$& $T_\mathrm{R}$&$\Delta T$\\ 
		\hline 
		$r_\mathrm{T,energy}$& 0.08 & -0.9753 & -0.0157 & 40 & 15 & 10\\ 
		\hline 
		$r_\mathrm{T,stage}$& 2.3823 & -0.6729 & -0.0329 & 40 & 15 & 15\\ 
		\hline 
		$r_\mathrm{T,egg}$&  0.511 & -0.2342 & -0.0824 & 40 & 20 & 2\\ 		
		\hline 
	\end{tabular} 
	\caption{Parameters of temperature influence model (Logan-10). $r_\mathrm{T,energy}$ influences energy consumption and life span, $r_\mathrm{T,stage}$ influences the transition rates between life stages, $r_\mathrm{T,egg}$ is used for influence of temperature on egg production.}
	\label{tab:parameters_T}
\end{table}

Bertinetti et al. \cite{Bertinetti2019} found that energy rich liquids (water, agar, milk) could prolong the life span of flies and increase egg production. 
The parameter values for each liquid and the extended equations for energy consumption and egg production can be found in Table~\ref{tab:parameters_other}. 

For the influence of lighting only white light and light hours per day are considered.
The effect of different wave lengths and light intensity is neglected.
It was found \cite{Hoc2019} that more light hours per day increase the amount of eggs harvested.
The reason however is not that female fertility is influenced by light, rather the chance of finding a partner for mating is increased \cite{Jones2021}.
Data from \cite{Hoc2019} was used to fit a model of the form
\begin{equation}
\label{eq:u_L}
r_\mathrm{L}(\tau) = a_1(1-\exp(-a_2\tau)).
\end{equation}
Parameter values are shown in Table~\ref{tab:parameters_other}.

\begin{table}[h]
	\centering
	\begin{tabular}{|c||c|c|c|}	
		\hline
		Feed& $k_\mathrm{fed1,f}$ & $k_\mathrm{fed1,m}$ & $k_\mathrm{fed2}$ \\ 
		\hline 
		none& 1.69 & 3.12 & 2.26  \\ 	
		\hline		
		water& 1.27 & 1.04 & 1.78  \\ 
		\hline 
		agar& 1 & 1 & 3.03 \\ 		
		\hline 
		milk& 0.58 & 0.87 & 4.06 \\ 	 
		\hline \hline
		Light& $a_1$ & $a_2$ &  \\ 
		\hline 
		& 1.8825 & 0.3711 &  \\  
		\hline 
	\end{tabular} 
	\caption{Parameter values for influence of different diets in breeding cage on energy consumption and egg production of BSF and parameters for impact of light on partner finding chance, see Eq.~\eqref{eq:u_L}.}
	\label{tab:parameters_other}
\end{table}

\subsection{Life sustainment on energy reserves}
From the diet in their larvae stage the flies have accumulated energy reserves $E_\mathrm{m}$ and $E_\mathrm{f}$.
In reality, starting reserves of each individual are statistically distributed so flies have varied life spans.
However, there is no data on energy reserves in BSF therefore the energy of each individual fly is assumed with a normalized starting value of 1. 
The starting values for the energy reserves of the populations $E_\mathrm{f,0}$ and $E_\mathrm{m,0}$ are the starting numbers of flies $N_\mathrm{f,0}$ and $N_\mathrm{m,0}$ respectively.
Note, that while a possible interpretation of $E$ is energy, it is hard to actually measure it in terms of Joule, hence it will be further referred to as energy but it is actually unitless.

\begin{figure}[h]
	\includegraphics[width=0.95\linewidth]{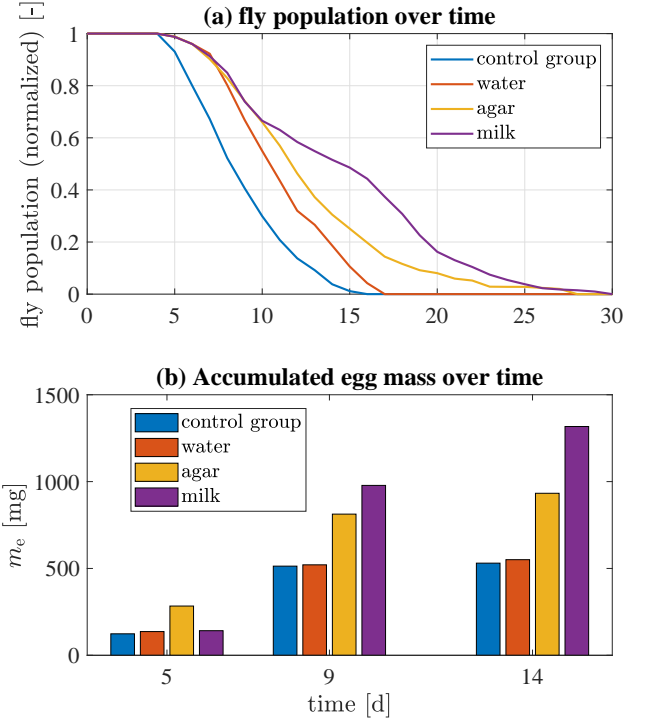}
	\caption{(a) BSF adult female fly colony size over time on different diets. Data for males looks similar. (b) Accumulated egg mass over time on different diets. Control group had no access to any energy/water source. Data is from \cite{Bertinetti2019}.}
	\label{fig:lifespan_diet}
\end{figure}

Due to their mouth parts, the flies can only consume liquids like water, nectar or milk \cite{Bruno2019}.
Doing so slows down decay of energy reserves.
Bertinetti et al. \cite{Bertinetti2019} found that agar and milk had positive influence on longevity of flies as well as on produced egg mass, see Fig.~\ref{fig:lifespan_diet}.

%

Energy reserves deplete in two ways, one is passive maintenance which is draining reserves regardless of what the fly does, the other are certain activities such as exploration to search for a partner.
The rate of change of energy reserve $\dot{E}$ can be described as
\begin{equation}\label{eq:e_dot_separate}
\dot{E}=-\beta - \gamma \mu
\end{equation} 
where $\beta$ is the daily energy required for the flies to live and the term $\gamma \mu$ describes how due to aging and stress more energy is consumed over time.
The term 
\begin{equation}\label{eq:mu}
\mu =  (1-E/E_0).
\end{equation}
describes aging processes and increases as energy reserves deplete.

Lupi et al. \cite{Lupi2019} tested the survival rates of male and female flies in isolation and in a shared cage and found that isolated flies live significantly longer.
This suggests that the mating process drains a lot of energy and should be a separate term from the basic daily energy drain.
Now Eq.\eqref{eq:e_dot_separate} is extended for the mating acts
\begin{align}\label{eq:e_dot_shared_f}
\dot{E}_\mathrm{f}&=-\beta_\mathrm{f} - \gamma_\mathrm{f} \mu_\mathrm{f} - \epsilon_\mathrm{f}\dot{m}_\mathrm{e}\\\label{eq:e_dot_shared_m}
\dot{E}_\mathrm{m}&=-\beta_\mathrm{m} - \gamma_\mathrm{m} \mu_\mathrm{m} - \epsilon_\mathrm{m} N_\mathrm{mate}
\end{align} 
$\epsilon_\mathrm{f}$ and $\epsilon_\mathrm{m}$ describe the energy costs for mating, $\dot{m}_\mathrm{e}$ is the mass of eggs being laid and $N_\mathrm{mate}$ is the fraction of the fly population that has recently mated. 
Parameter values may be found in Table~\ref{tab:parameters_k_x}.

Once energy reserves are depleted, the flies die.
The lower the reserves the higher the dying rate.
The term $\mu$, introduced in Eq.~\eqref{eq:mu}, is now used to describe decrease in population:
\begin{equation*}
\dot{N} =  -\mu N.
\end{equation*}
The following section further explores population dynamics.


\subsection{Fly life stages and population dynamics}
The four fly life stages are 'young', 'active', 'mated' and 'old'.
The male population is modeled separately from female population and females have an additional life stage 'fertilized' which is used to calculate egg production.

Pupae and recently hatched flies count towards the first stage which lasts until approximately two days after emergence, until the flies are able to mate.
Flies during this stage will hence be referred to as 'young' flies $N_\mathrm{y}$.
Young flies cannot yet mate and do not contribute to egg production \cite{Tomberlin2002b}.
The young fly stage is followed by the sexually 'active' stage $N_\mathrm{act}$.
However, only a fraction $N_\mathrm{mate}$ of the active flies does actually 'mate'.
There is no guarantee to immediately find a mating partner.
Note that while $N_\mathrm{y}$ and $N_\mathrm{act}$ have separate population for each sex, $N_\mathrm{mate}$ is a shared compartment for males and females, since flies are considered to only mate once.

After mating once, male flies do not further contribute to egg production process and are now considered old $N_\mathrm{m,old}$, while females become fertilized $N_\mathrm{fert}$.
Fertilized flies are able to lay eggs at a rate of
\begin{equation}\label{eq:dot_m_e_eggproduction_simple}
\dot{m}_\mathrm{e} =  k_\mathrm{ovi} N_\mathrm{fert},
\end{equation}
where $k_\mathrm{ovi}$ is a constant for egg production per fertilized fly.
After oviposition the females $N_\mathrm{fert}$ are considered old $N_\mathrm{f,old}$.

To describe the travel dynamics from one compartment to the next, transition rates $k$ are introduced.
These describe how many flies, male and female respectively, move from one stage to the next per time unit, e.g. 
\begin{equation*}
\dot{N}_\mathrm{y} =-k_1N_\mathrm{y} - \mu N_\mathrm{y}.
\end{equation*}
Parameter values for $k$ can be found in Table~\ref{tab:parameters_k_x}.

\begin{table}[h]
	\begin{tabular}{|c|c|c|c|c|c|}	
		\hline
		$k_1$& $k_2$ & $k_3$ & $k_4$ &$k_5$ & $k_\mathrm{ovi}$    \\ 
		\hline
		0.34 & 0.35  & 1.84  & 0.3   & 0.79 & 0.79 \\  
		\hline 		
		\hline 
		$\epsilon_\mathrm{f}$&$\epsilon_\mathrm{m}$  & $\beta_\mathrm{f}$ & $\beta_\mathrm{m}$ & $\gamma_\mathrm{f}$ & $\gamma_\mathrm{m}$\\ 
		\hline
		0.0287 & 0.0404 &  1.22$\cdot10^{-4}$ & 9.25$\cdot10^{-5}$ & 0.3513 & 0.1773\\  
		\hline		
	\end{tabular} 
	\caption{Parameter values for transition rates $k$ and energy consumption rates $\beta$, $\gamma$ and $\epsilon$ at reference temperature 25~$^\circ$C. Parameters were estimated from \cite{nakamura2016small}.}
	\label{tab:parameters_k_x}
\end{table}

\subsection{Combined Model}
First, the energy balance equations from Eq.~\eqref{eq:e_dot_shared_f} and \eqref{eq:e_dot_shared_m} are extended. 
Both the choice of feed $k_\mathrm{fed1}$ and the temperature in the breeding cage $u_\mathrm{T}$ affect the rate at which energy reserves are expended.
In addition, these two factors -- $k_\mathrm{fed2}$ for feed and $r_\mathrm{T,egg}$ for temperature -- also influence the amount of eggs laid, which is coupled with high reserves expenditure for female flies.
The energy equations are now:
\begin{align}
\label{eq:energy_modfied_f}
&\dot{E}_\mathrm{f}=-\dfrac{k_\mathrm{fed1,f}}{r_\mathrm{T,energy}} \left( \beta_\mathrm{f}+\gamma_\mathrm{f}\mu_\mathrm{f}\right) N_\mathrm{f} - \dfrac{\epsilon_\mathrm{f}}{k_\mathrm{fed2}r_\mathrm{T,egg}}\dot{m}_\mathrm{e}\\
\label{eq:energy_modfied_m}
&\dot{E}_\mathrm{m}=-\dfrac{k_\mathrm{fed1,m}}{r_\mathrm{T,energy}} \left( \beta_\mathrm{m}+\gamma_\mathrm{m}\mu_\mathrm{m}\right) N_\mathrm{m} - \epsilon_\mathrm{m}N_\mathrm{mate}.
\end{align}

Next, the population dynamics are modified.
Recall the fly life stages: young $N_\mathrm{y}$, active $N_\mathrm{act}$, old $N_\mathrm{old}$ as well as fertilized females  $N_\mathrm{fert}$ and the combined population $N_\mathrm{mate}$ of flies that have recently mated.
In all population dynamics there is a term $\mu N$ that shows the dying rate. 
Flies at any life stage can die.
Dying rate increases as the remaining reserves become less, i.e. with increasing time.
Transition rates $k$ are affected by temperature through $r_\mathrm{T,stage}$.
Additionally, transition rate from \textit{active} to \textit{mated} is also affected by light hours:
\begin{align}
	\label{eq:u_T,mate}
	k_i(T)&=k_j\cdot r_\mathrm{T,stage}(T)\quad j \in [1,\dots,5]\\
	\label{eq:u_T,tau,mate}
	k_3(T,\tau)&=k_3(T)\cdot r_\mathrm{L}(\tau).
\end{align}
A fly leaving \textit{young} stage is a loss for $N_\mathrm{y}$, but at the same time a gain for $N_\mathrm{act}$.
With this, the modified population models look a follows: 
\begin{align}
\label{eq:pop_modfied_young_f}
\dot{N}_\mathrm{f,y}&=  						 -k_1(T)N_\mathrm{f,y}												- \mu_\mathrm{f}N_\mathrm{f,y}\\
\label{eq:pop_modfied_young_m}
\dot{N}_\mathrm{m,y}&=  					 -k_2(T)N_\mathrm{m,y} 												- \mu_\mathrm{m}N_\mathrm{m,y}\\
\label{eq:pop_modfied_active_f}
\dot{N}_\mathrm{f,act}&=k_1(T)N_\mathrm{f,y} 	 -k_3(T,\tau)N_\mathrm{f,act}\frac{N_\mathrm{m,act}}{N_\mathrm{m,0}} 	- \mu_\mathrm{f}N_\mathrm{f,act}\\
\label{eq:pop_modfied_active_m}
\dot{N}_\mathrm{m,act}&=k_2(T)N_\mathrm{m,y} 	 -k_3(T,\tau)N_\mathrm{f,act}\frac{N_\mathrm{m,act}}{N_\mathrm{m,0}} 	- \mu_\mathrm{m}N_\mathrm{m,act}\\
\label{eq:pop_modfied_mate}
\dot{N}_\mathrm{mate}&=k_3(T,\tau)N_\mathrm{f,act}\frac{N_\mathrm{m,act}}{N_\mathrm{m,0}}  -k_4(T)N_\mathrm{mate}	- \mu_\mathrm{fm}N_\mathrm{mate}\\
\label{eq:pop_modfied_fert}
\dot{N}_\mathrm{fert}&=k_4(T)N_\mathrm{mate}  -k_5(T)N_\mathrm{fert}	- \mu_\mathrm{f}N_\mathrm{fert}\\
\label{eq:pop_modfied_old_f}
\dot{N}_\mathrm{f,old}&= k_5(T)N_\mathrm{fert} 			- \mu_\mathrm{f}N_\mathrm{f,old}\\
\label{eq:pop_modfied_old_m}
\dot{N}_\mathrm{m,old}&= k_4(T)N_\mathrm{mate} 	 		- \mu_\mathrm{m}N_\mathrm{m,old}
\end{align}

Lastly, the egg production is
\begin{equation}\label{eq:eggproduction_modified}
\dot{m}_\mathrm{e}=r_\mathrm{T,fed}(T)k_\mathrm{fed2}k_\mathrm{ovi}N_\mathrm{fert},
\end{equation}
where influence of temperature and feed was already explained at the beginning of the subsection.

Using this model, an optimal trajectory for control variables breeding cage temperature $T$ and light hours per day $\tau$ is calculated through optimzation algorithms introduced in the following section.

\section{Optimization of Egg Production}

The model described in the previous section was developed  with the purpose of manipulating and controlling egg production. 
This is done by manipulating the environmental factors temperature and light hours per day.
Equations \eqref{eq:u_logan_main} and \eqref{eq:u_L} show how the system is influenced by temperature $T$ and light hours $\tau$ respectively.
We assume an outside temperature of 20$^\circ \mathrm{C}$, i.e. not heating is require to hold 20$^\circ \mathrm{C}$ in the breeding cage and temperature below that requires cooling and causes costs.
The input is thereby defined as $u=\left[T \quad \tau\right]^\top \in \mathcal{R}^2$.
Lower and upper bounds for $T$ are chosen as 15~$^\circ$C and 40~$^\circ$C respectively under consideration of data from \cite{Chia2018}.
Upper bound for $\tau$ is 24~hours, lower bound is chosen as 2~hours as a minimum stimulus to keep flies alive \cite{Hoc2019}.

We now want to maximize egg output $m_\mathrm{e}$ with minimal control effort  i.e. low input $u$.
The accumulated number of eggs is considered as the output at the end of the process $m_\mathrm{e}(t_\infty)$.
However, it is also of interest to produce the eggs in shorter time so $m_\mathrm{e}$ during the process is incorporated as well.
While the negative of $m_\mathrm{e}$ is used in this minimization problem, the input $u$ is coupled as a positively signed term.
The optimization problem is formulated as follows
\begin{equation*}
\begin{split}
\min_{ u} \quad - Sm_\mathrm{e}(t_\infty) \int &-Qm_\mathrm{e}+u^\top R u\, \mathrm{d}t \\
\mathrm{s.t.}\quad &\eqref{eq:energy_modfied_f}, \eqref{eq:energy_modfied_m} \; \mathrm{and}\; \eqref{eq:pop_modfied_young_f} \;\mathrm{to} \; \eqref{eq:eggproduction_modified}\\		
& 15~^\circ\mathrm{C} \le T \le 40~^\circ\mathrm{C}\\
& 2~\mathrm{h} \le \tau \le 24~\mathrm{h}
\end{split}
\end{equation*}
with $Q$, $R$ and $S$ being the weights for egg mass, input and final egg mass respectively.

Time discretization is chosen as per day, because light hours $\tau$ can only be applied per day.
CasADi framework with Matlab is used for computation \cite{Andersson2019}.
Weights are chosen heuristically as $Q=10^4$, $R$ a diagonal matrix with $R_{11}=R_{22}=1$ and $S=10^5$ to obtain interpretable results.
They could also be adjusted to economic criteria.
In the next section optimization results are compared to a standard approach.

\section{Results}
\subsection{Parameter identification and model validation} 
Data from \cite{nakamura2016small} was used to estimate parameters from Table~\ref{tab:parameters_k_x}, \cite{Chia2018} for Table~\ref{tab:parameters_T} and \cite{Hoc2019} and \cite{Bertinetti2019} for Table~\ref{tab:parameters_other}.
Fly population and egg production were observed in similar experiments conducted by \cite{nakamura2016small} and \cite{Bertinetti2019}.
Their data is used for model validation.
Note that \cite{nakamura2016small} conducted their experiments at 25$^\circ \mathrm{C}$ while \cite{Bertinetti2019} at 28$^\circ \mathrm{C}$.
Also note that individual life stages could not be validated, since only data for fly population (male and female) and egg production is available.

\begin{figure}[h]
	\centering
	\includegraphics[width=0.98\linewidth]{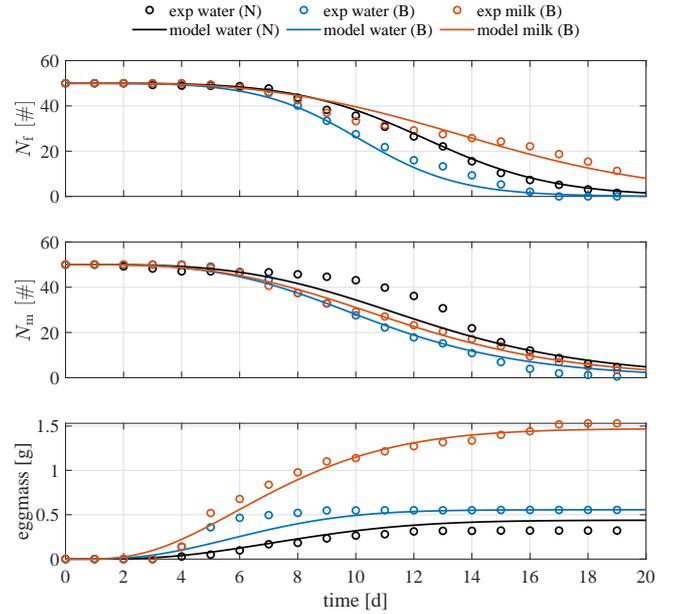}
	\caption{Model prediction versus experiment data of fly population and egg mass. Model explains data reasonably well.}
	\label{fig:modelvalidation}
\end{figure}

In Fig.~\ref{fig:modelvalidation}, model predictions versus data from literature is shown (\cite{nakamura2016small} control and \cite{Bertinetti2019} control and milk).
Simulations indicate that the model is able to predict states reasonably well.
Male population from \cite{nakamura2016small}(water -- black line) shows discrepancies around days 8 to 14.
This seems to be an outlier in data since flies should not live longer from water than milk as indicated in Fig.~\ref{fig:lifespan_diet}.
Egg mass is predicted well for Nakamura et al. \cite{nakamura2016small}(water -- black line) and milk from Bertinetti et al. \cite{Bertinetti2019} (milk -- red line).
There are, however, discrepancies to water (B), as the model predicts slower increase in egg mass than was measured.
A possible explanation is the difference in population genetics, flies in water (B) might have reached sexual maturity faster and can therefore produce eggs earlier.
Using the validated model, process optimization is performed.

\subsection{Results of optimal control} 
\begin{figure}[h]
	\centering
	\includegraphics[width=0.95\linewidth]{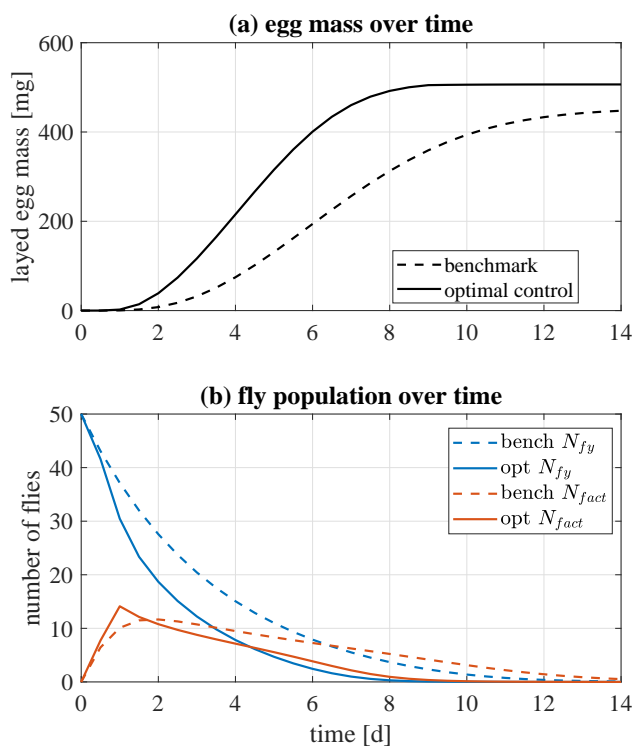}
	\caption{State trajectories for benchmark and optimal control. (a) With optimal control, more egg mass is accumulated in shorter time. (b) With optimal control, fly population declines faster but also reaches sexual maturity earlier and can produce eggs earlier.}
	\label{fig:bench_vs_oc_states}
\end{figure}

\begin{figure}[h]
	\centering
	\includegraphics[width=0.95\linewidth]{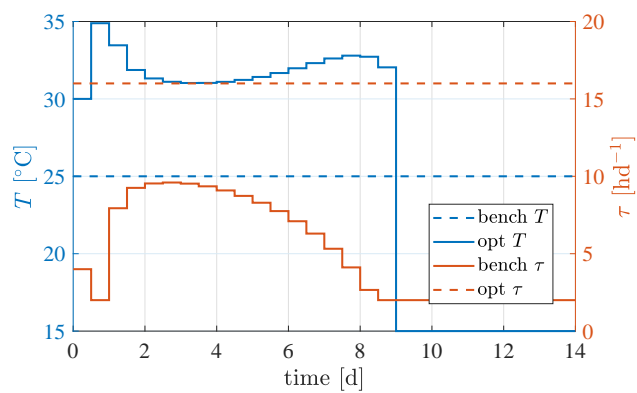}
	\caption{Input trajectories for benchmark optimal control. With optimal control more heating is needed until day 9. Also, with optimal control less light hours are applied overall to the system.}
	\label{fig:bench_vs_oc_inputs}
\end{figure}

The standard benchmark scenario is chosen similar to \cite{nakamura2016small}, with 16 hours of light per day and constant 25$^\circ$C with water nourishment.
Resulting state trajectories can be seen in Figure~\ref{fig:bench_vs_oc_states}, input trajectories can be seen in Figure~\ref{fig:bench_vs_oc_inputs}.

Comparison of states shows that, while fly population declines faster, the amount of eggs is significantly increased. 
The shorter life cycle results from faster transition between life stages due to higher temperature.
Flies reach sexual maturity faster and also burn their energy reserves faster. 
Shorter life cycles however mean that breeding cages can be refilled more often resulting in higher egg output.
400~mg of eggs can be collected after 6 days with optimal control, for the same amount more than 10 days are needed with benchmark conditions, i.e. a speed up by 40~\%.

Table~\ref{tab:results} shows comparison of four performance evaluation criteria.
Terms $\Sigma T$ and $\Sigma \tau$ are the sums of temperature (in degree days) and light hours respectively.
Input trajectories from optimization result in higher expenses for temperature and lower for lighting compared to benchmark, $\Sigma T$ is 5\% higher while $\Sigma \tau$ is 35\% lower.
Egg output is increased by 13\% and most flies are done mating by day 8 for optimization compared to day 12 for benchmark.

\begin{table}[h]
	\centering
	\begin{tabular}{|c|c|c|c|c|}	
		\hline
		& $\Sigma T$ & $\Sigma \tau$ & $N_\mathrm{fy,20\%}$ & $m_\mathrm{e}(t_\infty)$\\ 
		\hline 
		benchmark& 350~$^\circ$Cd & 224~h & 12~d  &447.6~mg\\ 	
		\hline		
		optimal control& 370~$^\circ$Cd & 72~h & 8~d  &506.5~mg\\ 
		\hline 
		
	\end{tabular} 
	\caption{Accumulated inputs, final egg mass and $N_\mathrm{fy,20\%}$ the time when female young population is less than 20\% of starting value.}
	\label{tab:results}
\end{table}



\section{Conclusion and Outlook}

A mechanistically motivated model for the adult life cycle of BSF was developed and presented in this work.
Parameters where obtained through fitting to data from literature.
Using this model, optimal trajectories for inputs temperature and light hours where calculated to maximize egg production at reasonable energy costs.
Results show an increase in egg mass production to 113\% while control effort was at 105\% and 65\% for heat and light respectively. 
This showcases the potential of model based process control in optimizing the egg production process with faster and higher production at reduced operation costs.

Dynamic models are in general useful tools for life cycle assessment and process monitoring, where comparison of measurements to model prediction may help in detecting possible faults.
Other uses are process analysis through simulation studies and --- as showcased in this work --- process optimization and control.
By adjusting the weights $Q, R$ and $S$ different priorities can be set, for either higher production or lower resource consumption.
Extensions for faster production cycles -- by penalizing time taken until $\dot{m}_\mathrm{e}$ falls below a threshold value -- can also be implemented.

Possible extensions of the model could include identification of more influence variables.
For that studies could be conducted on how for example fly activity during active stage may be affected.
If flies are in the air instead of idling in at cage walls the chance to find a mating partner is increased.
Sound, light flashes or air jets are promising examples for possible inputs.
Environmental factors such as temperature and humidity have an effect on egg eclosure and adult emergence -- processes which could also be included in the model.


%



\bibliographystyle{bib/IEEEtran}
\bibliography{bib/IEEEabrv,bib/bibliography}

\end{document}